\documentclass[twoside]{dis07}
\usepackage[latin1]{inputenc}
\usepackage{graphicx,epsfig,color}
\usepackage{wrapfig,rotating}
\usepackage{amssymb,amsmath,array}

\begin{document}
\title{Longitudinal Spin Measurements with Inclusive Hadrons in Polarized p+p Collisions at 200 GeV}

\author{Frank Simon (for the STAR Collaboration)
%
\vspace{.3cm}\\
Massachusetts Institute of Technology\\
77 Massachusetts Ave, Cambridge, MA 02139, USA
}

\maketitle
\begin{abstract}
We present measurements of the double longitudinal spin asymmetries for inclusive $\pi^0$ and $\pi^{+(-)}$ production in polarized p+p collisions at $\sqrt{s}$ = 200 GeV at mid-rapidity with the STAR detector from the 2005 RHIC run. These measurements are used to access $\Delta G/G$, the gluon polarization in the proton. The observed unpolarized inclusive cross sections show good agreement with NLO pQCD calculations. The double longitudinal spin asymmetries are compared to NLO pQCD calculations based on different assumptions for the gluon polarization in the nucleon to provide constraints on $\Delta G/G$. At the present level of statistics the measured asymmetries disfavor a large positive gluon polarization, but cannot yet distinguish between other scenarios.
\end{abstract}

\nocite{slides}
\section{Introduction}

A primary goal of the polarized p+p program at the Relativistic Heavy Ion Collider (RHIC) is the determination of the gluon polarization $\Delta G$ in the proton via spin asymmetry measurements in a variety of processes \cite{Bunce:2000uv}. Inclusive processes such as neutral pion, charged pion, and jet production provide sensitivity to gluons through the dominant subprocesses $gg \rightarrow gg$ and $qg \rightarrow qg$ at low and intermediate $p_T$. Because of the difference in the sign of the polarization for up and down valence quarks, the difference of the measured asymmetries in the $\pi^{+}$ channel and the $\pi^{-}$ channel is sensitive to the sign of the gluon polarization in the $p_T$ range where $qg \rightarrow qg$ contributions are sizable. These inclusive probes have only modest luminosity requirements and are natural first steps in longitudinal spin program. The unpolarized cross sections provide constraints on fragmentation functions and important validation of the NLO pQCD calculations used to interpret the measured spin asymmetries. The STAR experiment, with its large acceptance tracking and calorimetry, is uniquely capable of full jet reconstruction at RHIC, allowing a direct study of fragmentation through the association of a detected $\pi^0$ with its parent jet.

\section{Analysis and Results}

STAR detects neutral pions near mid-rapidity with its barrel electromagnetic calorimeter (BEMC) via reconstruction of the invariant mass of photon pairs; charged pions at high transverse momentum are identified in the time projection chamber (TPC) via their specific energy loss, $dE/dx$ \cite{Simon:2006xt}. For both these measurements, a trigger on the electromagnetic energy deposited in the calorimeter is crucial to select events with a hard initial scattering. For the present analyses two types of calorimeter triggers were used. The high tower (HT) trigger selects events with high energy deposition in one calorimeter tower ($\Delta \eta \times \Delta \varphi = 0.05 \times 0.05$), making this a good $\pi^0$ and photon trigger. The jet patch (JP) trigger selects events with significant energy deposited in a trigger patch of the calorimeter ($\Delta \eta \times \Delta \varphi = 1 \times 1$).  This is used to trigger on the electromagnetic energy in a jet. Each of these triggers is used with two different energy thresholds, referred to as HT1, HT2, JP1 and JP2. To achieve high detector live-time for the highest energy triggers, the minimum bias (MB), the HT1, and the JP1 triggers are prescaled during data taking. For the 2005 run period, only half of the BEMC was fully installed and commissioned, giving a coverage of $0 < \eta < 1$ for all azimuthal angles $\varphi$. This limits the acceptance for calorimeter triggers and for neutral pion reconstruction in this run.

The unpolarized cross sections for both neutral and charged pions have been discussed in more detail in  \cite{Simon:2006xt,Adams:2006nd}. Good agreement with NLO pQCD calculations \cite{Jager:2002xm} over several orders of magnitude in the cross section has been observed for both particle species. These measurements consistently favor the KKP fragmentation functions  \cite{Kniehl:2000hk}  over other sets.

\begin{figure}
\begin{minipage}[t]{.49\textwidth}
 \centering
  \includegraphics[width=.99\textwidth]{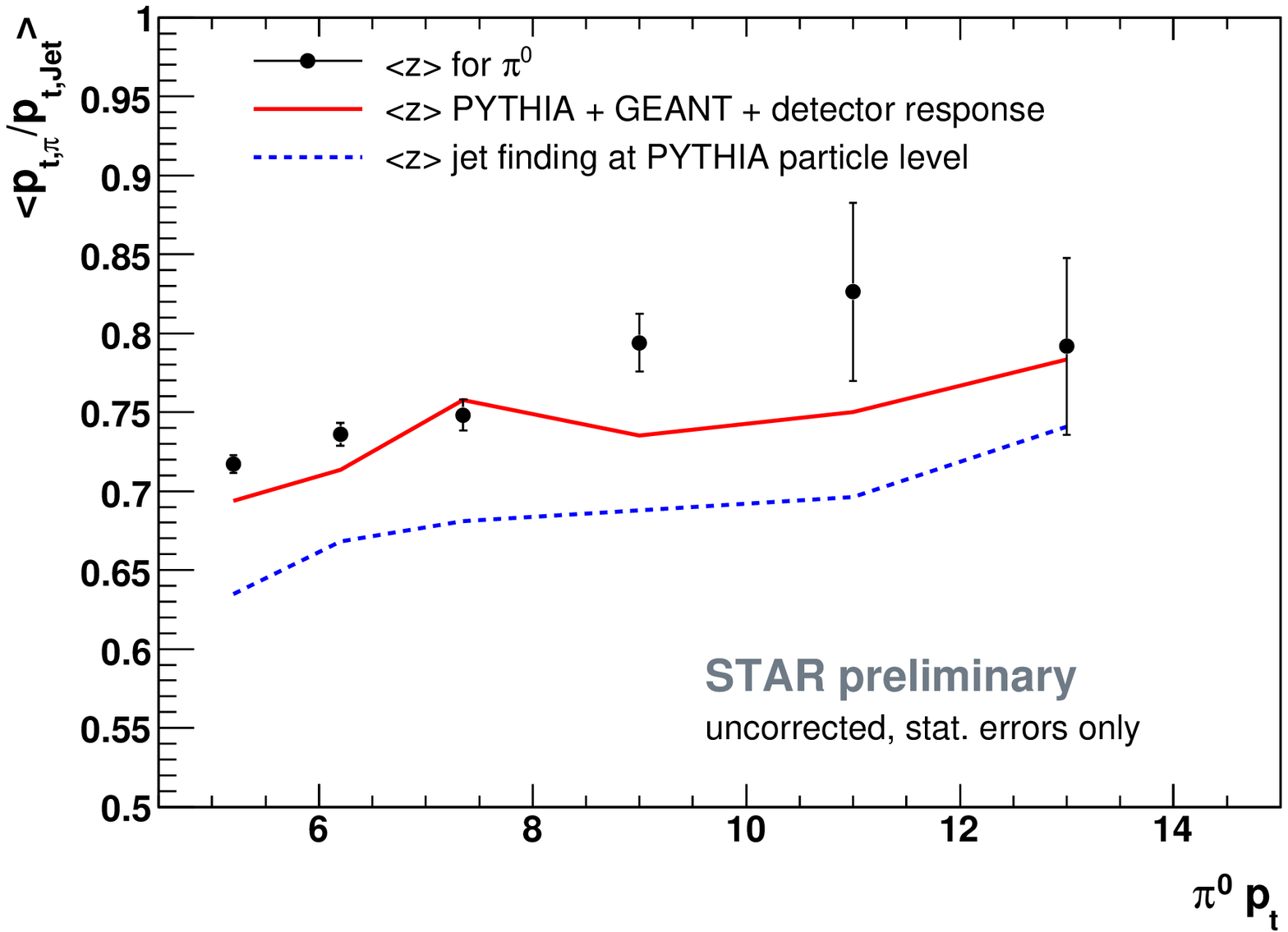}
  \end{minipage}
  \hfill
\begin{minipage}[t]{.49\textwidth}
  \centering
  \includegraphics[width=.99\textwidth]{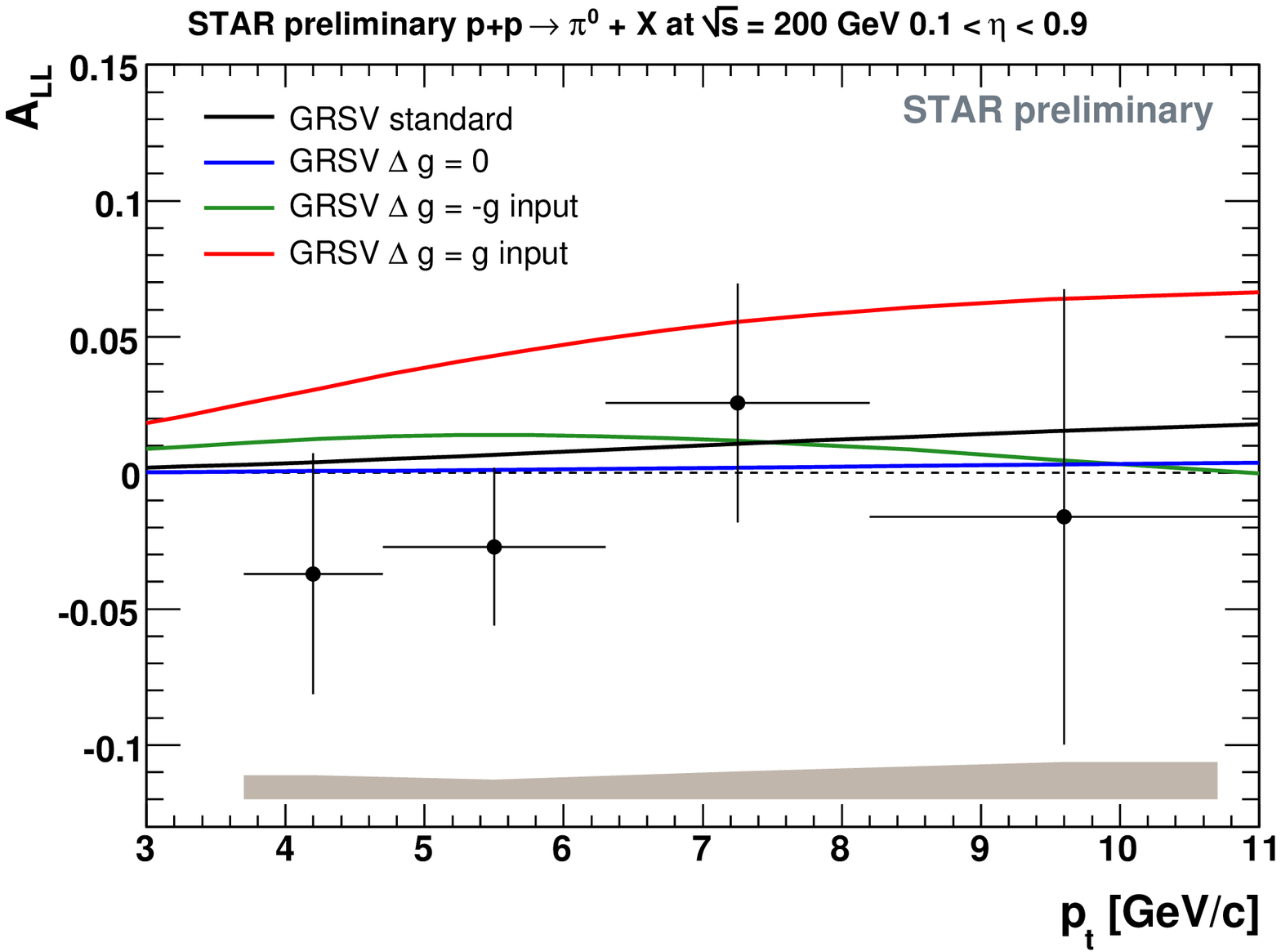}
    \label{fig:All}
 \end{minipage}
\caption{{\bf Left}: Mean momentum fraction of HT1 triggered $\pi^0$ in their associated jet as a function of $p_T$. The data points are plotted at the bin center in $p_T$ and not corrected for acceptance or trigger effects. Only statistical errors are shown. The solid red and the dashed blue line show the $\pi^0$ $\left<z\right>$ for PYTHIA simulations with a full GEANT detector simulation and with jet finding on the PYTHIA particle level, respectively, indicating the size of resolution and reconstruction effects.\newline
{\bf Right}: Double longitudinal spin asymmetry for inclusive $\pi^0$ production together with NLO pQCD predictions based on different assumptions for $\Delta G$. The systematic error shown by the gray band does not include a 9.4\% normalization uncertainty due to the polarization measurement.}
  \label{fig:MeanZ} 
 \end{figure} 	

To investigate the momentum fraction carried by high $p_T$ $\pi^0$'s in their parent jet, identified neutral pions were associated with jets found \cite{Abelev:2006uq} in the same event. An association was made if the pion was within the jet cone of 0.4 in $\eta$ and $\varphi$. To avoid edge effects, the analysis is restricted to $0.4 < \eta_{jet} < 0.6$. Because other analyses have shown that calorimeter-only jets without any charged tracks are often associated with beam background, a maximum ratio of the neutral to total energy of 0.95 is imposed. The left side of figure \ref{fig:MeanZ} shows the mean momentum fraction $\left<z\right>$ of neutral pions associated with jets for HT1 triggers as a function of the pion's $p_T$. The momentum fraction is not corrected for acceptance, efficiency or resolution of the jet reconstruction. The mean momentum fraction of $\pi^0$ in electromagnetically triggered jets is approximately 0.75, and rises slightly with $p_T$, consistent with measurements of leading charged hadrons in jets in fixed-target experiments \cite{Boca:1990rh}. Also shown in this figure is the mean momentum fraction of neutral pions in jets in PYTHIA \cite{Sjostrand:2000wi} simulations, both with jet finding on the PYTHIA particle level, and with a full detector simulation in GEANT. The difference between these two gives an impression of the size of the detector and reconstruction effects on the mean momentum fraction. These effects are dominated by the jet reconstruction, since the momentum determination for neutral pions in the electromagnetic calorimeter is very precise. The simulation results suggest that resolution and reconstruction effects increase the observed $\left<z\right>$ by about 10\%.

The longitudinal double spin asymmetry is given by
\vspace{-2mm}
\begin{equation}
A_{LL} = \frac{1}{P_1 P_2}\frac{(N^{++} - RN^{+-})}{(N^{++} + RN^{+-})},
\vspace{-2mm}
\end{equation}
where $P_{1,2}$ are the mean measured beam polarizations and $R$ is the ratio of integrated luminosities for equal and opposite beam helicities. $N^{++}$ and $N^{+-}$ are the particle yields in equal and opposite beam helicity configurations, respectively. The polarizations are obtained with the RHIC polarimeters \cite{Okada:2006dd}. Typical polarization values during the run period were $\sim$ 50\%. The relative luminosities are monitored in STAR with the BBCs. Typical $R$ values were around 1.1. The integrated luminosity for the asymmetries for both neutral and charged pion production is $\sim$ 1.6 pb$^{-1}$ from the 2005 RHIC longitudinally polarized proton run.

\begin{figure}
\begin{minipage}[t]{.49\textwidth}
	\centering
	\includegraphics[width=.99\textwidth]{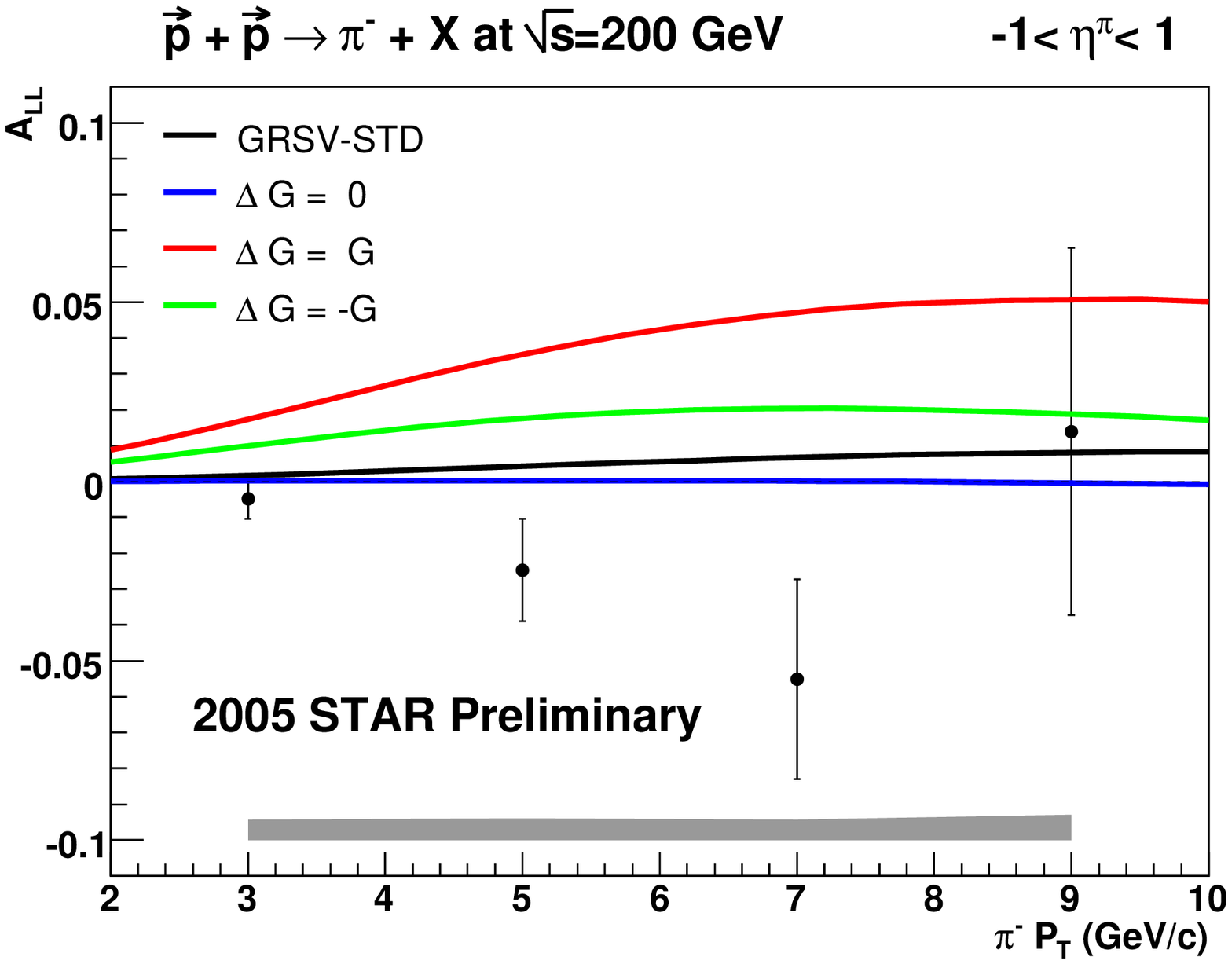}
	\label{fig:all_plus}
\end{minipage}
\hfill
\begin{minipage}[t]{.49\textwidth}
	\centering
	\includegraphics[width=.99\textwidth]{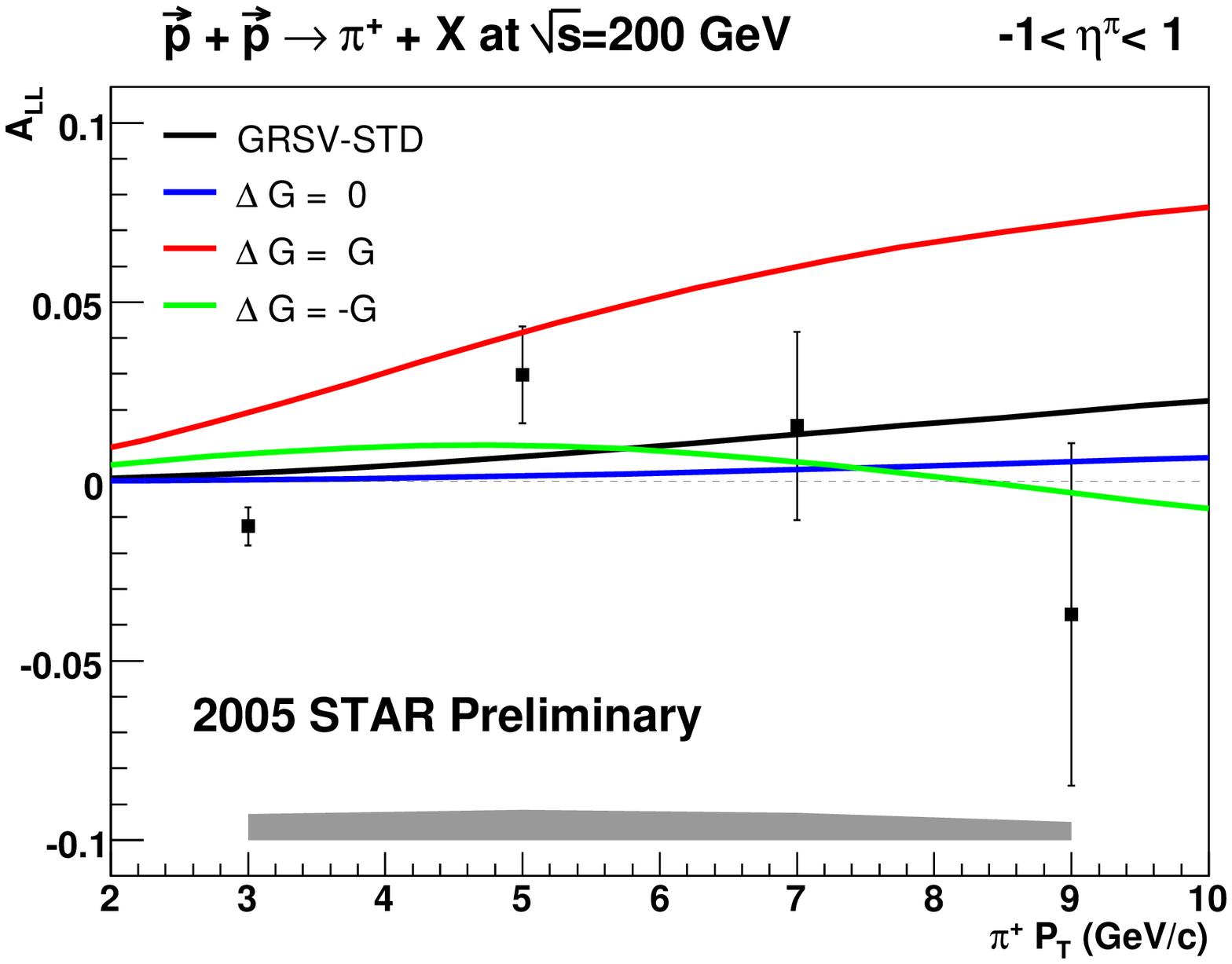}
	\label{fig:all_minus}
\end{minipage}
\caption{Double longitudinal spin asymmetries for inclusive charged pion production.  $A_{LL}(\pi^{-})$ is displayed in the left panel and $A_{LL}(\pi^{+})$ is on the right.  The asymmetries are compared to theoretical predictions of $A_{LL}$ incorporating various scenarios for the gluon polarization.  The error bars are statistical; point-to-point systematic uncertainties are added in quadrature and shown as the gray band at the bottom of each figure.  A scale uncertainty of 9.4\%  from the uncertainty on the beam polarization measurements is not included.}
\label{fig:all}
\end{figure}

The right side of figure \ref{fig:MeanZ}  shows the measured double spin asymmetry for inclusive $\pi^0$ production, together with theoretical predictions assuming different gluon polarization scenarios \cite{Jager:2004jh}. The systematic errors shown in the figure include contributions from $\pi^0$ yield extraction and background subtraction, remaining background, possible non-longitudinal spin contributions and relative luminosity uncertainties. An overall normalization uncertainty of 9.4\%, due to errors on the polarization values obtained with the RHIC polarimeters, is not included. Studies of parity-violating single spin asymmetries and randomized spin patterns show no evidence for bunch to bunch or fill to fill systematics. The GRSV standard curve is based on the best fit to DIS data; the other curves show scenarios of extreme positive, negative and vanishing gluon polarizations. The data are consistent with three of these evaluations and tend to disfavor the scenario with a large positive gluon polarization.

Figure \ref{fig:all} shows the double longitudinal spin asymmetries for inclusive charged pions \cite{Kocoloski:2006he}. The same theoretical predictions as for the $\pi^0$ case are also shown here. To obtain a flavor-separated result the KKP fragmentation functions \cite{Kniehl:2000hk} used for the calculations are modified by multiplying favored fragmentation functions by $(1+z)$ and unfavored ones by $(1-z)$. The leading systematic error in this analysis is the trigger bias, because the events are triggered on electromagnetic energy deposition, while charged hadrons are analyzed. This is different from the $\pi^0$ case, where  the trigger is directly sensitive to neutral pions. As in the case of the $\pi^0$ asymmetry, the results are least consistent with the maximal positive gluon polarization scenario, but discerning among the other scenarios is limited by statistics. The double longitudinal asymmetry of ($\pi^+ + \pi^-$) has been found to be consistent with the asymmetry of $\pi^0$.

\section{Conclusion}

The STAR experiment at the Relativistic Heavy Ion Collider has obtained preliminary results on the cross section and the double longitudinal spin asymmetry of inclusive neutral and charged pion production in polarized p+p collisions at $\sqrt{s}$ = 200 GeV. The measured cross sections are found to be in good agreement with NLO pQCD predictions. The longitudinal asymmetries disfavor the maximal positive gluon polarization scenario, but currently have no resolving power among other scenarios due to limited statistics. With the large increase in sampled luminosity and polarization in the 2006 run, a significant improvement of the statistical power of these analyses is expected in the near future.


\begin{footnotesize}

\end{footnotesize}



\begin{thebibliography}{99}
\bibitem{slides}
Slides:
  {\texttt{http://indico.cern.ch/contribution{D}isplay.py?contrib{I}d=143\&session{I}d=4\&conf{I}d=9499}}.

\bibitem{Bunce:2000uv}
  G.~Bunce, N.~Saito, J.~Soffer and W.~Vogelsang,
  Ann.\ Rev.\ Nucl.\ Part.\ Sci.\  {\bf 50}, 525 (2000).

\bibitem{Simon:2006xt}
  F.~Simon  [STAR Collaboration],
  AIP Conf.\ Proc.\  {\bf 870}, 428 (2006)
  [arXiv:hep-ex/0608050].
  
   \bibitem{Adams:2006nd}
  J.~Adams {\it et al.}  [STAR Collaboration],
  Phys.\ Lett.\  B {\bf 637}, 161 (2006).


\bibitem{Jager:2002xm}
  B.~Jager, A.~Schafer, M.~Stratmann and W.~Vogelsang,
  Phys.\ Rev.\  D {\bf 67}, 054005 (2003).


\bibitem{Kniehl:2000hk}
  B.~A.~Kniehl, G.~Kramer and B.~Potter,
  Nucl.\ Phys.\  B {\bf 597}, 337 (2001).

  
 \bibitem{Abelev:2006uq}
  B.~I.~Abelev {\it et al.}  [STAR Collaboration],
  Phys.\ Rev.\ Lett.\  {\bf 97}, 252001 (2006). 

\bibitem{Boca:1990rh}
  G.~Boca {\it et al.},
  Z.\ Phys.\  C {\bf 49}, 543 (1991).

\bibitem{Sjostrand:2000wi}
  T.~Sjostrand {\it et al.},
  Comput.\ Phys.\ Commun.\  {\bf 135}, 238 (2001).

\bibitem{Okada:2006dd}
  H.~Okada {\it et al.},
  arXiv:hep-ex/0601001.

\bibitem{Jager:2004jh}
  B.~Jager, M.~Stratmann and W.~Vogelsang,
  Phys.\ Rev.\  D {\bf 70}, 034010 (2004).

\bibitem{Kocoloski:2006he}
  A.~Kocoloski,
  arXiv:hep-ex/0612005.

\end{thebibliography}
\end{document}